\documentclass[iop,revtex4]{emulateapj}

\usepackage{amsmath,amssymb}
\usepackage[breaklinks,colorlinks,urlcolor=blue,citecolor=blue,linkcolor=blue]{hyperref}
\usepackage{epsfig}    
\usepackage{graphicx}    
\usepackage{lineno}
\usepackage{natbib}
\usepackage{bigints}
\usepackage[outdir=./]{epstopdf}
\usepackage{longtable}
\usepackage{lineno}
\usepackage{lscape}

\newcommand\teff{\ensuremath{T_\text{eff}}}

\newcommand{\rearth}{R$_{\oplus}$}
\newcommand{\msun}{M$_{\odot}$}
\newcommand{\rsun}{R$_{\odot}$}

\newcommand{\kep}{{\it Kepler}}

\newcommand{\kicnum}{KIC~6423922}
\newcommand{\koinum}{KOI~6705.01}

\providecommand{\adsurl}[1]{\href{#1}{ADS}}

\slugcomment{Accepted to ApJ 16 November 2015}

\shorttitle{Enigmatic and Ephemeral KOI~67051}
\shortauthors{Gaidos, Mann \& Ansdell}

\begin{document}

\title{The Enigmatic and Ephemeral M Dwarf System KOI 6705: Cheshire Cat or Wild Goose?}

\author{Eric Gaidos \altaffilmark{1,2,3}} 
\affil{Department of Geology \& Geophysics, University of Hawai`i at M\={a}noa, Honolulu, HI 96822}
\email{gaidos@hawaii.edu}
\author{Andrew W. Mann \altaffilmark{1,4}}
\affil{Department of Astronomy, University of Texas at Austin, Austin, TX 78712}
\and
\author{Megan Ansdell}
\affil{Institute for Astronomy, University of Hawai`i at M\={a}noa, Honolulu, HI 96822}

\altaffiltext{1}{Visiting Astronomer at the Infrared Telescope Facility, which is operated by the University of Hawaii under contract NNH14CK55B with the National Aeronautics and Space Administration.}
\altaffiltext{2}{Visiting Scientist, Observatoire de Sauverny, Universite de Gen\`{e}ve}
\altaffiltext{3}{Visiting Scientist, Center for Astrophysics, Harvard University}
\altaffiltext{4}{Harlan J. Smith Postdoctoral Fellow, Department of Astronomy, University of Texas at Austin}

\begin{abstract}

We confirm a 0.995~d periodic planetary transit-like signal, \koinum, in the \kep{} lightcurve of the star \kicnum. Optical and infrared spectra show that this star is a mid M-type dwarf with an effective temperature $= 3327 \pm 60$K, metallicity [Fe/H] $= -0.08 \pm 0.10$, radius $= 0.31 \pm 0.03$\rsun{}, and mass $= 0.28 \pm 0.05$\msun{}. The star is $\approx 70$~pc away and its space motion, rotation period, and lack of H$\alpha$ emission indicate it is an older member of the "thin disk" population.  On the other hand, the star exhibits excess infrared emission suggesting a dust disk more typical of a very young star.  If the \koinum\ signal is produced by a planet, the transit depth of 60 ppm means its radius is only $0.26^{+0.034}_{-0.029}$\rearth{}, or about the size of the Moon.  However, the duration ($\gtrsim 3$~hr) and time variation of \koinum\ are anomalous: the signal was undetected in the first two years of the mission and increased through the latter two years.  These characteristics require implausible orbits and material properties for any planet and rule out such an explanation, although a dust cloud is possible.  We excluded several false positive scenarios including background stars, scattered light from stars that are nearby on the sky, and electronic cross-talk between detector readout channels.  We find the most likely explanation to be that \koinum\ is a false positive created by charge transfer inefficiency in a detector column on which \kicnum\ and a 1.99~d eclipsing binary both happened to fall. 

\end{abstract}

\keywords{stars:low-mass -- stars:fundamental parameters -- planets and satellites:formation -- methods:spectroscopic}

\section{Introduction}

By one definition, our Solar System contains only eight planets, but the Sun's entourage includes smaller "dwarf" planets such as Pluto and Ceres, considerably more numerous asteroids and comets, and still smaller bodies down to dust grains. Surveys of other stars have revealed more size diversity among exoplanets, including objects more massive than Jupiter, "super-Earths" with radii intermediate Earth and Uranus \citep{Haghighipour2013}, and planets smaller than Earth \citep{Sinukoff2013}. A full understanding of planet formation and evolution is not possible without a complete and accurate picture of the planet gamut. Studies of very small objects are particularly important because canonical planet formation theory predicts that rocky Earth- and super-Earth-size planets accrete from smaller planetary "embryos" \citep{Morbidelli2012}. Although these predecessors are usually lost to incorporation, some may be preserved in a state of arrested planetary development:  Mars may be one such body \citep{Dauphas2011}. Searches for smaller objects around other stars may also reveal a variety of phenomena that are also pieces of the planetary puzzle, including "exocomets" \citep{Lecavelier1999}, "exomoons" \citep{Kipping2015}, and "disintegrating" planets \citep{Rappaport2012}.      

Earth-size and smaller planets are difficult to detect. Ironically, among the first planets to be discovered was a Mercury-sized planet orbiting the pulsar PSR B1257+12 \citep{Wolszczan1994}. But no other "pulsar planet" systems has been uncovered, and instead it is the unprecedented accuracy and continuity of photometry obtained by the NASA \kep{} mission that has enabled exploration of this realm \citep{Borucki2010}. The smallest planet confirmed to date is \kep{}-37b, with a radius of only $0.30 \pm 0.06$\rearth{} \citep{Barclay2013}. 

Planets are detected in \kep{} data if and when they transit their host stars, and the signal is proportional to the square of the ratio of the planet radius to the stellar radius. All else being equal, it is easier to find smaller planets around smaller stars, i.e. M dwarfs. The prime \kep{} mission observed several thousand M dwarfs and a disproportionate number of Earth-size and smaller (candidate) planets have been found identified around these low-mass stars \citep{Gaidos2013,Dressing2015}, including a Mars-size planet \citep[\kep{}-42d,][]{Muirhead2012} and a similar-size candidate planet that is on a 4.2~hr orbit near its Roche limit \citep{Rappaport2013}.  On the other hand, analysis and interpretation of weak signals at the threshold of \kep{} detection must be wary of numerous astrophysical false positives and subtle instrumental artifacts \citep[e.g.,][]{Santerne2013,Coughlin2014}.

Gaidos et al. (MNRAS, submitted) revisited the parameters of M dwarfs observed in the \kep{} prime mission and the properties of their planets detected thus far, using the DR24 release of \kep{} Objects of Interest (KOIs, Coughlin et al., in prep.) as the source catalog.  One object, \koinum, stands out as having an estimated size about that of the Moon, and thus among the smallest planets detected to date.  Here, we describe our follow-up observations and analysis of this intriguing object.  Our ground-based spectroscopy and imaging are described in Sec. \ref{sec.obs}. In Sec. \ref{sec.star} we determined the properties of the star, and our false-positive probability analysis is contained in Sec. \ref{sec.fp}.  Our independent re-analysis of the \kep{} data to estimate transit and planet parameters is in Sec. \ref{sec.planet} and in Sec. \ref{sec.discussion} we discuss the apparent nature of \koinum\ that emerged from our analysis.

\section{Observations}
\label{sec.obs}

\subsection{Spectroscopy}

\kicnum\ was observed with the Super-Nova Integral Field Spectrograph (SNIFS) on the UH 2.2m telescope on Maunakea during the night of UT 25 June 2015. We also obtained SNIFS spectra of two nearby stars (KIC 6423914 and 6423941) on UT 2 October 2015. Using a lenslet array, SNIFS re-images a target onto gratings in separate blue (3200-5200\AA) and red (5100-8700\AA) channels with a resolution $R \approx 900$  \citep{Aldering2002,Lantz2004}. We obtained a signal-to-noise ratio (SNR) $\approx 100$ for each star. Spectra were extracted and wavelength calibrated using flat-fields and arcs taken at the same pointing.  The reduction and flux calibration involved observations of spectrophotometric standards and an airmass correction based on data accumulated over several years of observations: details are given in \citet{Mann2012} and \citet{Lepine2013}.

We obtained a near-infrared ($JHK$) spectrum of \kicnum\ on UT 7 July 2015 with the uSpeX spectrograph \citep{Rayner2003} on the NASA Infrared Telescope Facility on Maunakea.  The spectrum was taken in short cross-dispersed mode with a 0.3" slit to obtain maximum resolution ($R \approx 2000$). The SpeXTool pipeline \citep{Cushing2004} was used to perform de-biasing, flat-fielding, extraction, and wavelength calibration. Spectra of bright A0 stars were used in the Xtellcor routine \citep{Vacca2003} to calibrate fluxes and remove telluric lines. We refer the reader to \citet{Mann2015} for more details on the procedures. 

A higher-resolution ($R \approx 8000$) optical spectrum of the star was obtained with the Echellette Spectrograph and Imager \citep[ESI,][]{Sheinis2002} on the Keck 2 telescope on Maunakea on UT 17 July 2015. The spectrum was obtained in cross-dispered mode with a 0.5" slit and the peak SNR was $\approx 120$. A white dwarf standard (EG 131) was used as a spectrophotometric standard. Images were processed and the spectra extracted using the ESIRedux package\footnote{http://www2.keck.hawaii.edu/inst/esi/ESIRedux/index.html} \citep{2003ApJS..147..227P,2009PASP..121.1409B}. 

\subsection{Adaptive Optics Imaging}

Adaptive optics (AO) images of \kicnum\ were taken with the Keck-2 telescope using the laser guide star system on UT 1 October 2015. Observing strategy followed that of Kraus et al. (submitted). Observations were done with the $K$-prime pass-band and the facility AO imager, NIRC-2, in vertical angle mode. All observations used the smallest pixel scale (9.952 mas pix$^{-1}$). Observations consisted of 8 images, each 20s, four at each of two dither positions. 

For each AO image we corrected for geometric distortion using the NIRC-2 distortion solution from \citet{Yelda2010}, flagged dead pixels, and removed cosmic rays. We fit the star in each of the eight images with a simple point-spread function built from stacking other single stars taken in the same night. We measured the standard deviation of the fluxes among all 5-pixel annuli around the primary star, effectively identifiying any aperture with a $\geq 5\sigma$ outlier as an astrophysical source. We stacked the limits from the eight images and used this as our detection limit for a given projected separation and contrast associated with that $5\sigma$ value. 

\subsection{Archival Data}

We retrieved long-cadence \kep{} Pre-search Data Conditioning Simple Aperture Photometry \citep[PDCSAP,][]{Stumpe2012} lightcurves of KOI~6705 for 12 \kep{} observing quarter from the NASA Exoplanet Archive. The star fell on defunct CCD module 3 during \kep{} observing quarters 1, 5, 9, 13, and 17. A digitized image from the first Palomar Optical Sky Survey (POSS-I) centered at the location of \kicnum\ was obtained through the Space Telescope Science Institute archive. We obtained a corresponding extract of a high-resolution $J$-band image from the WFCAM Science Archive \citep{Hambly2008}. This image was obtained as part of an infrared survey of the \kep{} field with the UKIRT Wide-Field Camera \citep[WFCAM,][]{Casali2007} in good seeing ($\sim 0.8$").  We retrieved photometry from the Wide-field Infrared Survey Explorer \citep[WISE,]{Wright2010} as compiled in the AllWISE source catalog.  The closest source in the catalog, J185657.47+414907.6, is located 2.4" to the SW of the 2MASS position of \kicnum, and is detected as a single, isolated source in all 4 bands.  The epochs of the 2MASS and WISE observations are 1998.39 and 2010.56, and the expected proper motion is 2.5", also to the SW.  Thus the WISE source is thus at the exact position of \kicnum\ at the time of observation.  The star is detected in all four (W1-W4) bands (3.4, 4.6, 12 and 22 $\mu$m) with magnitudes $11.191 \pm 0.022$, $11.048 \pm 0.021$, $10.473 \pm 0.061$, and $8.581 \pm 0.23$.

\begin{figure}
\begin{center}
\includegraphics[width=3.25in]{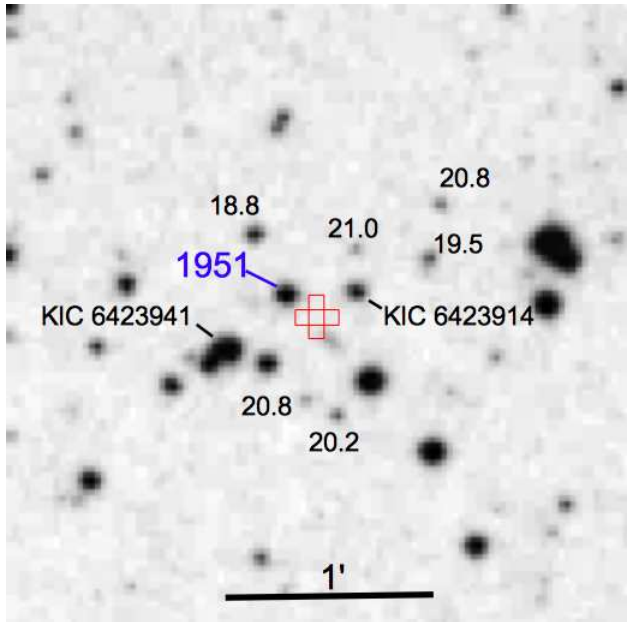}
\caption{Three arc-min square image from a digitized Palomar Optical Sky Survey (POSS)-I plate (epoch 1951.67) centered on the J2000 coordinates of \kicnum, showing the proper motion of the star over 60 years. North is up and east is to the left. A representation of the 5-pixel aperture mask used by the DR24 pipeline is placed at the approximate 2011 location of the star. The USNO B1 magnitudes of several of faint stars are given; there is no comparable background star at the present location of \kicnum. The two labeled KIC stars are discussed in the text.\label{fig.dss}}
\end{center}
\end{figure}

\begin{figure}
\begin{center}
\includegraphics[width=\columnwidth]{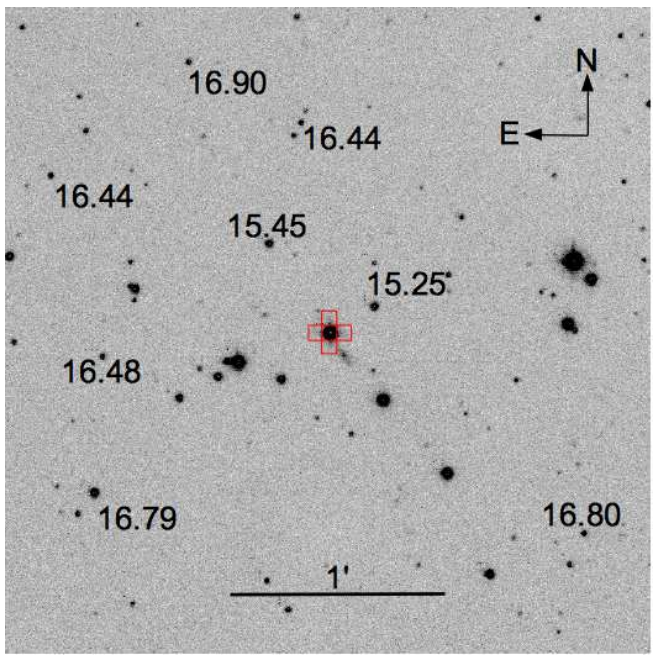}
\caption{Three arc-min square image from the UKIRT $J$-band survey of the \kep{} field (epoch 2009.6) centered on \kicnum. Position, scale, and orientation are the same as for Fig. \ref{fig.dss}. The 2MASS $J$-band magnitudes of several faint stars are given.\label{fig.ukirt}}
\end{center}
\end{figure}

\section{Stellar Parameters}
\label{sec.star}

No emission in the Balmer H$\alpha$ line was found in either the SNIFS or ESI spectra of \kicnum. A decimal spectral type was assigned based on the strength of the CaH and TiO features in the SNIFS optical spectrum and the empirical relations from \citet{Lepine2013}. Effective temperature \teff{} was estimated by comparing the SNIFS optical spectrum to a grid of PHOENIX model spectra constrained by metallicity (see below). This comparison was calibrated with a set of stars with accurately measured angular radii and bolometric fluxes and temperatures established using the Stefan-Boltzmann relation \citep{Mann2013}. We used the CFIST grid of models constructed with the \citet{Caffau2011} relative abundances for the Sun. Metallicity was derived from measurements of metal-sensitive features in the SpeX NIR spectrum following the procedure described in \citet{Mann2012}. We find that \kicnum\ is a main-sequence M dwarf with a decimal spectral type of 3.4, a \teff{}=$3327\pm60$~K, and [Fe/H]$=-0.08\pm0.10$ ([M/H] =$-0.08\pm0.09$). 
Radius, mass, and luminosity were computed using \teff{}, metallicity, and empirical relations derived by observations of a set of nearby calibrator M dwarfs with established properties \citep{Mann2015}. We find that the star has a radius of $0.305 \pm 0.030$\rsun{}, a mass of $0.277 \pm 0.048$\msun{}, and an absolute $K$-band magnitude $M_K = 7.05 \pm 0.30$.  

\begin{figure*}[t]
\begin{center}
\includegraphics[width=\textwidth]{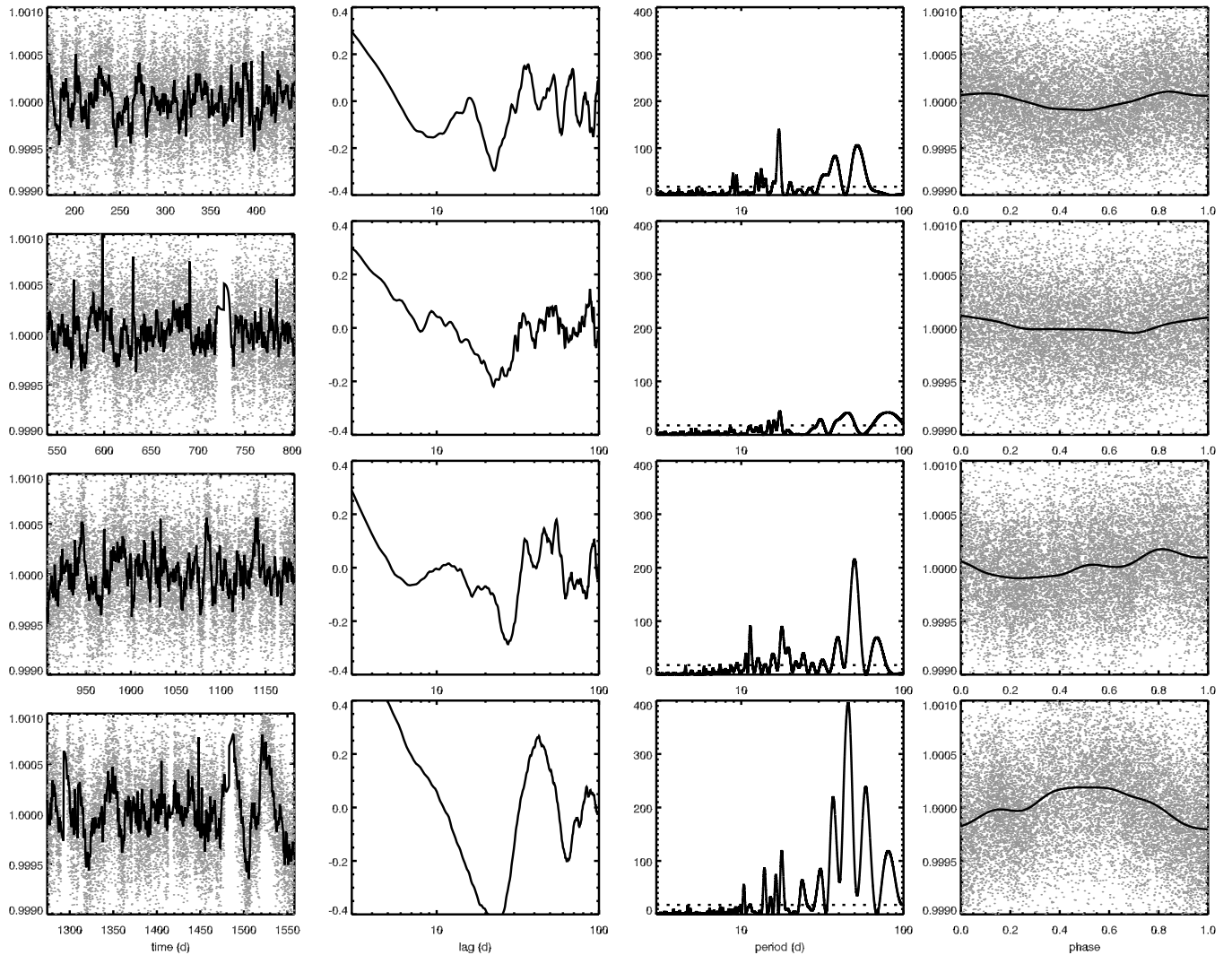}
\caption{Analysis of four segments of \kep{} time-series photometry of \kicnum, each spanning 9 months, in temporal sequence from top row to bottom row. {\it Far left:} PDCSAP lightcurve plus a Gaussian-smoothed version with $\sigma$ equal to 10 times the cadence. {\it Middle left:} Auto-correlation function of the smooth lightcurve. {\it Middle right:} Lomb-Scargle periodogram of the original lightcurve. The dash line marks the threshold for a false alarm probability of $3 \times ^{-3}$. {\it Far right:} Lightcurve phased with the period of the strongest signal in the periodogram.}
\label{fig.rotation}
\end{center}
\end{figure*}

We analyzed the \kep{} lightcurve of \kicnum\ to search for rotational variability and determine the rotation period, a qualitative indicator of stellar age. Because of the 90-day gaps when the star was on the defunct CCD module, we analyzed each set of three contiguous quarters separately (Fig. \ref{fig.rotation}). We calculated the auto-correlation function (ACF) of a version of the lightcurve smoothed with a Gaussian having $\sigma$ equal to 10 times the cadence (5 hr). We also calculated a Lomb-Scargle periodogram \citep{Scargle1982} over 3-100~d.

The spectral properties of the lightcurves vary;  in both Q2-4 and Q6-8 the strongest peak in the periodogram is near 17~d. However this is not reflected in the ACF, with only a weaker peak for Q2-4. In Q10-12 and Q14-16 these are replaced by stronger periodogram peaks at 50~d and 46~d, respectively, and there is an equivalent peak at about 47~d in the ACF for Q10-12 (but 35~d for Q14-16). Changes in the power spectrum of rotational variability can be explained by migration of multiple groups of star spots, or by variations in the activity of multiple stars. (See \citet{Walkowicz2013} or \citet{Reinhold2013} for a discussion of this effect). We interpret the weaker 17~d period as a 3:1 harmonic of a $\sim50$-d rotation period produced by multiple spot groups; these coalesce or decrease to a single spot group after Q9. \citet{McQuillan2013} found the distribution of rotation periods in \kep{} M dwarfs to be bimodal, with peaks near 17 and 33~d.\footnote{They did not analyze the lightcurve of \kicnum.}  They found that proper motions of the 33~d rotators are statistically higher, suggesting they belong to an older population of stars. 

The space motion of a star is another clue to its origin and age. We estimated the radial velocity of \kicnum\ by correlating the ESI spectrum with synthetic spectra generated by the PHOENIX stellar atmosphere model with the \citet{Caffau2011} relative abundances, as described above, and using the offset (in $\log$ wavelength) to determine the redshift. A heliocentric correction was then applied to arrive at -3.2 km~sec$^{-1}$. Proper motions of PM$_{\alpha} = -0.154$~arcsec~yr$^{-1}$ and PM$_{\delta} = -0.136$~arcsec~yr$^{-1}$ were obtained from the SUPERBLINK catalog \citep[][S. L\'{e}pine, private communication]{Lepine2005,Lepine2011}. We estimated the absolute magnitude $M_K = 7.05 \pm 0.26$ of the star using the empirical relations from \citet{Mann2015}. Since, $K_s = 11.38$, the estimated distance is $74 \pm 10$ pc. Combined with the proper motions this yielded a $(U,V,W)$ space motion of (-65.6, -18.2, +36.5) km~sec$^{-1}$ with respect to the Local Standard of Rest estimate of \citet{Coskunoglu2011}. On a Toomre diagram, $\sqrt{U^+W^2} = 75$~km~sec$^{-1}$ and \kicnum\ falls within the thin disk population defined by \citet{Fuhrmann2004} but close to the boundary with the thick disk population. This suggests an age at least several Gyr and is consistent with the lack of H$\alpha$ emission and a slightly sub-solar metallicity.

Paradoxically, WISE observations of \kicnum\ suggest excess 12 and 24$\mu$m emission suggestive of circumstellar dust and characteristic of much younger stars.  The 2MASS-WISE colors are $K_s-W3 = 0.91 \pm 0.06$ and $K_s - W4 = 2.80 \pm 0.23$, with a marginally significant excess ($K_s - W2 = 0.34 \pm 0.03$ at $4.6\mu$m.  These put this star amongst the "evolved disk" population of $\approx 10$Myr-old Upper Scorpius stars in a color-color diagram \citep[Fig. 2 in ][]{Luhman2012}.  Evolved disks are depleted, optically thin versions of primordial circumstellar disks, without the central hole characteristic of "transition" disks. 

The angular resolution of the WISE survey in the W3 and W4 bands is 6" and 12", respectively, and thus source confusion is a possibility.  The nearest (7") source in the DSS images and 2MASS Point Source Catalog is a spiral galaxy the (see Section \ref{sec.proximal}).  The galaxy is not detected as a separate WISE source, is too faint for 2MASS and it falls outside the SDSS footprint, but we associated it with KIC~6423919, which has $r = 17.44$ and $g-r = 0.77$.  To correct for interstellar reddening/extinction we use the mean reddening value $E_{B-V} = 0.079$ at this location from \citet{Schlafly2011} and adopting $R=A/E(B-V)$ of 3.31 for $g$ and 2.32 for $g$ from \citet{Yuan2013}, yielding a dereddened $g-r = 0.69$ and an un-extincted $r = 17.26$.  This $g-r$ color is typical of "normal" galaxies.  

To estimate the WISE magnitudes of this galaxy we identified analogs from the atlas of galaxies with well-characterized spectral energy distributions and integrated WISE fluxes of \citet{Brown2014}.  We selected 15 galaxies with de-reddened $0.66 < g-r < 0.72$ of which we retained 8 classified as spirals that are not obviously interacting with other galaxies or in groups.  The de-reddened $r-W3$ and $r-W4$ colors of these galaxies are correlated and span several magnitudes.  The two most infrared luminous cases are the LIRG/Seyfert galaxies Z~453-62 and NGC~7591 ($r-W3 \approx 2.1$ and $r-W4 = 3.2$).  If these values provide an upper limit for KIC~6423919, then W3$>15.3$ and W$>14.2$, and the galaxy does not significantly contribute to the observed WISE emission of \kicnum.  

\section{False-Positive Analysis}
\label{sec.fp}

\subsection{Proximal Sources}
\label{sec.proximal}

We estimated the false-positive probability (FPP) that the transit signal is not associated with the star \kicnum\ but is instead a periodic but non-transit signal from an unresolved or undetected background star.\footnote{This is different from the possibility that the signal is from an unresolved companion, which we address in Sec. \ref{sec.discussion}.}  We did not assume that the signal had to be produced by an eclipsing binary (EB), although this is a popular scenario.  We also did not factor in the probability that the background star could actually produce the transit signal, i.e. be an eclipsing binary, thus this calculation is conservative. 

The false positive calculation used a prior based on a model of the stellar population in the field, and likelihoods based on observational constraints. We used the TRILEGAL model (version 1.6) \citep{Vanhollebeke2009} to calculate the stellar population to $K_p = 27$ occupying three square degrees centered at the position of \kicnum\ (392314 stars). This magnitude limit exceeds the faintest stars that could possibly produce the a signal with the amplitude of \koinum\ (i.e. complete occulting). We randomly selected stars from this population and placed them uniformly in a circular field of 15" radius centered on \kicnum. To fully account for uncertainties in stellar and transit parameters we randomly sampled the error distributions (or the posterior distributions from the transit fits) with each Monte Carlo iteration.   

We applied six likelihood factors: (1) the background star can produce the observed transit depth; (2) the predicted motion of the image centroid during the transit is consistent with \kep{} observations; (3) the transit duration is consistent with the density of the background star; (4) the star would not be visible at that location in the 60 year-old DSS image (Fig. \ref{fig.dss}); (5) the star would not appear in the vicinity of \kicnum\ in a $J$-band UKIRTF image (Fig. \ref{fig.ukirt}); and (6) the star would not appear in our Keck-2 NIRC-2 AO image in $K'$-band. 

The first factor is the likelihood that the contrast at the \kep{} pass-band is $\Delta K_p < -2.5 \log (\delta/R)$, where $\delta$ is the transit depth and $R$ is the ratio of the fraction of the background stellar flux entering the aperture to the fraction of the target star flux entering the aperture. \kicnum\ has $K_p = 15.67$ and thus any background star capable of producing the transit signal must be brighter than $K_p = 26.1$ (for the case of $R = 1$). To calculate $R$ we performed bilinear interpolations on the pixel response function for the appropriate detector channel (33, 49 or 77) using the tables provided in the Supplement to the \kep{} Instrument Handbook (E. Van Cleve \& D. A. Caldwell, KSCI-19033). 

For the second factor we calculated the probability that a transit of the model background star would have a duration $T$. For this we calculated the cumulative probability distribution of the quantity
$\Delta = T/(P^{1/3}\tau^{2/3})$, following \citet{Silburt2015},
\begin{equation}
\int n(\Delta) d\Delta =  \int_{0}^{1} \eta(e) de \int_0^{2\pi} d\omega \sqrt{1 - \frac{\Delta^2 \left(1 + e \cos \omega \right)^2}{1-e^2}},
\end{equation}
where $P$ is the period, $\tau = \sqrt{3/(G\rho_*}/\pi$ is the stellar free-fall time, $\rho_*$ is the mean stellar density, $e$ is the orbital eccentricity, and $\omega$ is the argument of periastron \citep{Gaidos2013}. The eccentricities of short-period ($<20$~d) binary stars tends to be small \citep{Duchene2013}. We described the eccentricity distribution as a Rayleigh function with a mean of 0.1.  To estimate the probability $p$ that the signal is associated with each background star, we calculated the likelihood $n(\Delta)$ for the background star and compared to the likelihood $n(\Delta_*)$ for \kicnum\ itself: 
\begin{equation}
p = \langle \frac{n(\Delta)}{n(\Delta) + n(\Delta_*)} \rangle,
\end{equation}
where the average is over the distribution of possible values of $T$. 

The centroid motion factor was calculated as the ratio of the posterior probability that the predicted effect of a background source is consistent with the observed motion to the total posterior probabilities for the target or background source scenarios. This reduces to:
\begin{equation}
p =  \left(1 + \exp \left[\frac{\left(x - \bar{x}\right)^2-x^2}{2\sigma_x^2} + \frac{\left(y- \bar{y}\right)^2-y^2}{2\sigma_y^2}\right]\right)^{-1},
\end{equation}
where ($x$,$y$) and ($\sigma_x$,$\sigma_y$) are the observed centroid motion and its uncertainties, and ($\bar{x}$,$\bar{y}$) is the predicted centroid motion due to a background source. For this particular case there are no other resolved sources in the \kep{} aperture postage stamp so the predicted centroid motion for the no-background scenario is zero. We adopted the centroid motion from the DR24 release: $-0.411 \pm 1.35$" in R.A., and $-0.381 \pm 0.636$" in declination.

There are no stellar-like sources in the 1951 DSS image that fall within the photometric aperture for \koinum\ (Fig. \ref{fig.dss}). We quantified this limit as $m > 21$ by comparing the DSS image to sources in the USNO-B catalog.  A disk galaxy (discussed in Sec. \ref{sec.star}) is centered $\approx 10$" from \kicnum\ and may partially intrude into the photometric aperture. Based on a $g = 18$ and a typical disk galaxy $M_g \sim -21$ \citep{Brown2014}, the distance modulus must be $\approx 39$ and the most massive, luminous stars must be fainter than $m=27$.  Even the total occulting of one of these stars would produce a signal smaller than 10 ppm, thus we do not consider this object a plausible source for the signal of 6705.01. 

Besides \kicnum, there are no other discernible sources in the UKIRT $J$-band image appearing within the \kep{} aperture. To approximately calibrate the image, we identified sources in the UKIRT image which have $J$-band photometry in the 2MASS Point Source Catalog. We found that sources with $J < 18$ can be ruled out further than 3" from \kicnum. No neighboring sources were identified in our Keck-2 NIRC-2 AO images. The $5\sigma$ detection limiting contrast is $\Delta K = 5.6$ at 0.15" separation and increases to 8.6 by 1.5" after which it is constant until the edge of the field at 2". Values for each Monte Carlo source were interpolated from a finite set of points. The NIRC-2 imaging offers no constraints on sources closer than 0.15" or further than 2".

We performed a running average of the FPP and monitored the value for convergence. After $\sim 10,000$ iterations the FPP had converged to $3 \times 10^{-5}$. This does not account for the probability that a background star has a companion on an eclipsing orbit, and thus the actual FPP is smaller. The only possible scenario involves a star that is outside the 2" FOV of NIRC-2 but still inside the \kep{} photometric aperture and is too faint for UKIRT $J>18$ or DSS ($K_p > 21$) but bright enough to cause the signal $K_p < 26$. The most likely scenario is a distant ($> 1$~kpc) M dwarf: other stars will be too bright.  

We estimated the rate of M dwarf EBs (MDEBs) using a log-normal distribution for orbital period $P$ with $\bar{a} = 5.3$~AU and $\sigma_P = 1.3$ \citep{Duchene2013}.  We normalized the distribution using a determination that $\sim$3.5\% of M dwarfs are close ($<0.4$ AU) binaries \citep{Clark2012}. The apastron-averaged probability of transit is $R_*/\left(a(1-e^2)\right)$. We integrated over a thermal distribution of eccentricities $\eta(e)de = 2 e de$ up to contact orbits where $e_{\rm max} = 1 - R_*/a$. The eclipse probability becomes
\begin{equation}
p_{\rm ecl}(P) = -\sqrt[3]{\frac{3\pi}{G \rho_* P^2}}\ln \left[1-\left(1-\left(\frac{3\pi}{G \rho_* P^2}\right)^{2/3}\right)\right]. 
\end{equation}
This is convolved with the period distribution to obtain an integrated MDEB probability. For a radius of 0.3\rsun{} and a total system mass of 0.6\msun{}, $p_{\rm ecl} \approx 8.5 \times 10^{-3}$.  This value is intermediate the EB statistics from \kep{} of 0.4\% for M dwarfs \citep{Shan2015} and 1.4\% for all stars \citep{Slawson2011}.  The {\it prior} probability that there is a background MDEB behind \kicnum\ becomes $3 \times 10^{-8}$ but since we selected this star for its periodic variability the posterior probability will be higher.

\subsection{Distal Sources}
\label{sec.distal}

False positives can also be produced by spurious signals from distal but bright, periodically variable stars, including EBs. \citet{Coughlin2014} identified four mechanisms which can yield such FPs: (a) scattered light in the point response function extending at least 50", and further for stars brighter than $K_p = 16$;  (b) cross-talk between the four readout channels on the same 2-CCD module, which can extend up to two degrees; (c) antipodal reflection around the optical axis of \kep{} which can produces a faint "ghost" on the opposite side of the \kep{} field of view; and (d) contamination along a CCD column from sources at lower-numbered rows due to charge transfer inefficiency (J. Coughlin, private communication), and thus acting at distances of up to 68'.  In all of these scenarios both the period and phase of the real stellar signal ("parent") and artifacts ("children") are preserved. 

We screened for these effects first by comparing the period and BKJD epoch of \koinum\ to the most recent DR24 catalog of Threshold Crossing Events (TCEs). We screened TCEs that had nearly identical periods to \koinum\ as well as its 2:1 super- and subharmonics. The only object (TCE 008257115-01) that has the same period within errors has an epoch that differs by 0.915 d. It is approximately $11^{\circ}$ from \kicnum\ and does not fall in the same CCD module or is even in the same region of the CCD for all four rotations.   Not all stars in the \kep{} field were observed by \kep{} and unobserved stars could also be the source of the signal. To identify candidate parents of a spurious signal we considered the full \kep{} Input Catalog \citep[KIC, ][]{Brown2011}. This includes stars as faint as 21st magnitude but is nearly complete to $K_p \approx 20$, judging from the distribution with magnitude compared to TRILEGAL predictions.

\begin{figure}[h]
\begin{center}
\includegraphics[width=\columnwidth]{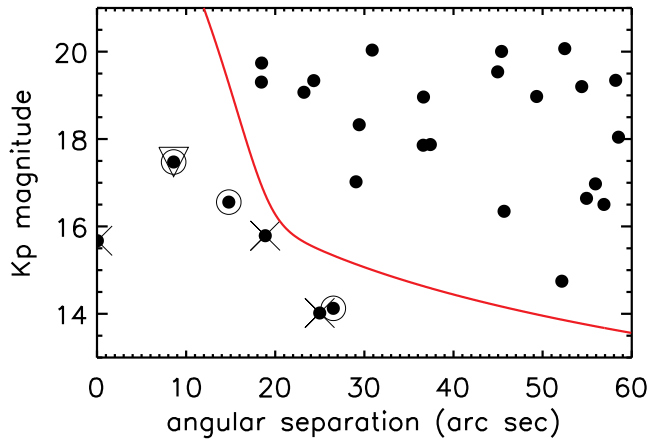}
\caption{\kep{} $K_P$ magnitude vs. angular separation for KIC stars within one arc minute of \kicnum. The red curve is the magnitude criterion for production of a false positive via contaminating scattered light in the point response function from an EB as formulated by Eqn. 9 in \citet{Coughlin2014} and assuming an eclipse depth of 50\%. There are five sources brighter than this limit, but two (crossed) of them are observed by \kep{} and do not exhibit variability. The inverted triangle is a galaxy.\label{fig.nearby}}
\end{center}
\end{figure}

We examined nearby stars on the sky that might produce a false positive via scattered light. Figure \ref{fig.nearby} plots the angular separation and $K_p$ magnitude of KIC sources within one arc minute of \kicnum\ (that star is also shown on ordinate). The red curve is the magnitude of an eclipsing binary with a depth of 50\% (the maximum possible) and a given angular separation that could induce a  \koinum-like signal via scattered light in the point response function \citep[Eqn. 9 in ][]{Coughlin2014}. The five sources below that curve all satisfy that criterion. We inspected the DSS image to verify that no other relevant sources were missed by the KIC. We also examined stars at larger separations (not shown) but found no additional candidates. 

Two of the five stars (crossed) were observed by \kep{} and do not appear in the TCE catalog; they are thus ruled out as possible parents of the \koinum\ signal. A third (inverted triangle) is the distant galaxy mentioned earlier. The remaining two are KIC~6423914 and 6423941, located 14" NW and 23" SE, respectively, from the present position of \kicnum\ (Fig. \ref{fig.dss}). \citet{Brown2011} classified these as a \teff{}$\approx 5400$~K G-type dwarf and a \teff{}$\approx 4600$~K K-type giant. We compared SNIFS optical spectra of these stars (Sec. \ref{sec.obs}) with spectra we obtained with the same instrument of a set of G and K dwarfs and giants with empirically-determined \teff{} \citep[][ and references therein]{Huang2015}. We confirmed the luminosity class assignments and estimated \teff{} as 5350~K and $<4700$K, respectively.  (The latter is an upper limit because our comparison sample lacks cooler giants).  

Based on the \citet{Coughlin2014} description of the \kep{} PRF and $K_p = 16.6$, the eclipse depth for KIC 6423914 would have to be $\sim 1.5$\%, which for a 0.8\rsun{} star corresponds to a Jupiter-size planet, brown dwarf, or very late M dwarf. But the density of the G-type primary is much greater than the lightcurve-based value ($\sim 0.1$, Fig. \ref{fig.nopriors}) and thus a highly eccentric orbit is still required. Such an orbit is implausible because tidal dissipation would have quickly circularized it. If KIC~6423941 is the EB it cannot have a radius of 10\rsun as estimated by \citet{Brown2011} because a 1~d orbit would lie inside the star.  It can be no larger than $\sim 3$\rsun, characteristic of stars climbing the red giant branch. Because KIC~6423941 ($K_p = 14.1$) is further away on the sky, the eclipse must be also deeper, $\sim 10$\%, which could only be produced if it was occulting a hotter dwarf companion. 

Because of the gradient in the PRF across the photometric aperture of KOI~6705, fluctuations in the scattered light from an EB would induce shifts in the apparent centroid of KIC~6423922. We calculated the centroid of this scattered light relative to the center using a finite-element method and compared this to the measured offsets and uncertainties from the \kep{} transit difference image. The predicted values are ($\alpha$, $\delta$) = (-2.76",1.22") and (1.07",-0.32") for KIC~6423914 and 6423941, respectively.  Comparing this to the measured offsets and uncertainties in the transit (difference) signal (Sec. \ref{sec.fp}), the significance of these offsets in standard deviations would be (1.7,2.5) and (1.1,0.1). Thus we can provisionally exclude KIC~6423914, but not 6423941. 

Sufficiently bright stars may have antipodal reflections ("ghosts") at the location opposite the center of the \kep{} field of view. If the bright star is variable then this can give rise to weak artifacts in the lightcurves of stars within 50" of the antipode \citep{Coughlin2014}. These "ghosts" are predicted to be 8.5 magnitudes fainter than the source star \citep{Caldwell2010}. The signal of \koinum\ could be produced by a star as faint as $K_p = 17$, depending on the angular separation from the antipode, if it was an EB with a maximum possible eclipse depth of 50\%. We examined the field within 50" of the antipodal location of \kicnum\ ($\alpha = 297.09342$, $\delta = 46.62356$). Only a single star, KIC~9844861, with $K_p = 14.6$ and a separation of 30", satisfies the requirements.\footnote{A second star, USNO-B 1366-0338191 has a USNO B-magnitude of 16.5 but it lies 49" from the antipode and is unlikely to be the source of the signal.}  This star was classified as a main-sequence F-type dwarf (\teff{} = $6362 \pm 149$K) by \citet{Huber2014} and was observed in all 17 \kep{} quarters. It does not appear in the DR24 catalog of TCEs and a Lamb-Scargle analysis of the lightcurve shows no signal near 0.995~d.  We thus exclude antipodal reflection as an explanation for the signal of \koinum.

We investigated the cross-talk and column anomaly scenarios \citep{Coughlin2014} by identifying all stars in the KIC catalog that, for any of the four observing seasons/rotations, fall within one CCD row and column of \kicnum and in the same CCD module (cross-talk), or fall within one column on the same CCD and at lower, positive row numbers (column anomaly). \citet{Coughlin2014} find the column anomaly to involve KOIs as faint as $K_p \approx 15$, and involve parents as much as $\sim 1.5$ magnitudes fainter than the KOI so we only considered stars as faint as $K_p = 17$. No stars that were identified had TCEs and we excluded stars that were observed by \kep{} but had no TCEs. We found no KIC stars that satisfied the cross-talk criteria but we identified 9 stars with $K_p < 17$ that fall at lower, positive row number and within one column of \kicnum\ during at least one observing season and are thus potential parents of the \koinum\ signal via column anomaly. This figure is not anomalous: based on TRILEGAL output, the density of stars with $K_p < 17$ at this pointing is 1.17 per sq. arc. min. and the expected number of stars falling in a box 3 columns wide and 770 rows long (the approximate location of \kicnum) is 12. 

\begin{figure}[h]
\begin{center}
\includegraphics[width=\columnwidth]{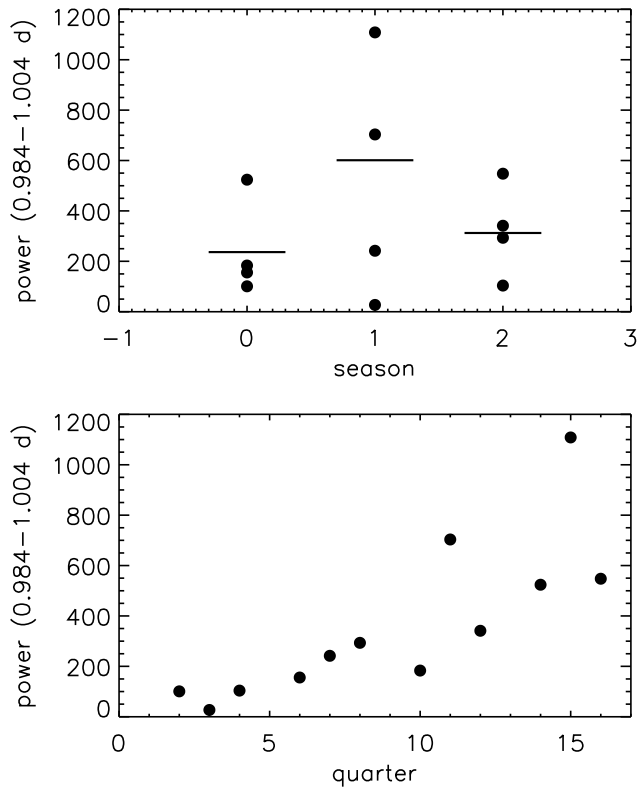}
\caption{{\it Top:} Total power in the period interval of 0.984-1.004~d in a Lomb-Scargle power spectrum of the KOI~6705 lightcurve for each quarter, grouped by observing season. The horizontal bars are the values from the power spectrum of the concatenated lightcurves for all quarters in a season. {\it Bottom:} Total power in the same interval plotted vs. sequential quarter showing the significant increasing trend. \label{fig.power}}
\end{center}
\end{figure}

We narrowed the list of potential sources of contamination by examining the dependence on observing season. Lomb-Scargle power spectra of the data from individual quarters, grouped by observing season, show that the \koinum\ signal, i.e. the total power at periods between 0.984 and 1.004~d, is higher in observing season 1 compared to 0 and 2 (top panel of Fig. \ref{fig.power}). However the scatter between quarters is large and the likelihood that the two quarters with the highest signal would occur in the same season (as in this case) by chance is 1/3. The point-biserial correlation coefficient of the mean power between season 1 and seasons 0 + 2 is 0.36; the probability that this coefficient would exceed this under the null hypothesis (no correlation) was calculated by Student's t-test to be $p = 0.13$ (10 degrees of freedom), and therefore the dependence on season is not statistically significant.  Instead, the \koinum\ signal shows a clear trend with observing quarter (bottom panel of Fig. \ref{fig.power}), with a Spearman rank coefficient of 0.92 ($p = 3 \times 10^{-5}$.  (We discuss the time variation in the  signal in Sec. \ref{sec.planet}).  The signal is clearly detected in the last three quarters, representing all three of the observable seasons.   Four of the 9 potential sources of a column anomaly (KIC~6503164, 6503213, 6586268, and 6758789)  fall on the same or a neighboring column of \kicnum\ in all three observing seasons.  

To test whether any of these stars or KIC~6423941 (discussed above) might be an EB with a period of 0.995~d (or some multiple of it) we identified two \kep{} Full Frame Images (FFIs) that were obtained within a few minutes of a predicted transit center time and compared these with corresponding images obtained in the same quarter (Q10 and Q15) but several hours outside of transit.  We shifted and differenced these images and inspected the signal at the location of each of the five stars.  As a check, we also identified a third Q10 FFI, also obtained out-of-transit, and differenced the two-out-transit images.  In both Q10 and Q15, KIC~6503213 exhibits a $\approx$30\% decrease in signal in the in-transit relative to the out-of-transit images, but negligible difference between any of the out-of-transit images.  None of the other four suspects exhibit significant variation.  The Q10 and Q15 events  are separated by an odd multiple of 0.995~d, which means that successive eclipses are of roughly equal depth and the binary is composed of similar stars with an orbital period of 1.99~d.   KIC~6503213 is classified in the KIC as an evolved late G- or early K-type star (\teff{} = 5164K) with a radius of 2.5\rsun{}.  Eclipses by such a star could reasonably be of 3-4 hr duration.  

\section{Transit and Planet Parameters}
\label{sec.planet}

We extracted a subset of the \kep{} PDCSAP data during the predicted transits along with a 3~hr buffer on each side. Data with obvious non-transit artifacts (e.g., stellar flares) and data covering less than half of a transit were identified by eye and removed. We fit the out-of-transit lightcurve with a third order polynomial to remove trends in the lightcurve unrelated to the transit (mostly from stellar variability), and then stacked the data into a single lightcurve. 

\begin{figure}[]
\begin{center}
\includegraphics[width=\columnwidth]{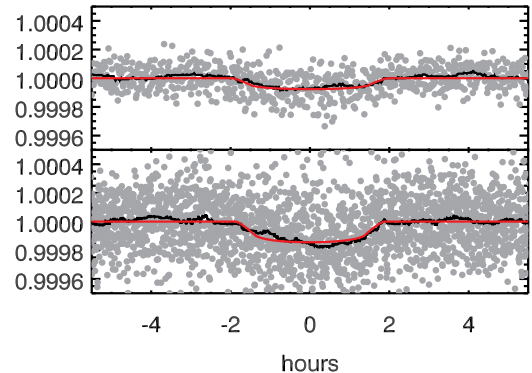}
\caption{Phased KOI~6705 lightcurve data for all quarters (top) and the last 600 d (bottom). In the top panel only, each point represents the average of 10 binned measurements. The black line is a running median with a width of 45 min (the expected transit duration if $b=0$) and the red line is a best-fit \citet{MandelAgol2002} model.  The top panel includes all the data and the best-fit model corresponds to the parameter reported in Table \ref{tab.params}, while the bottom panel uses only data from the last 600~d of observations when the signal is strongest. \label{fig.phased}}
\end{center}
\end{figure}

We determined the period of the \koinum\ signal to be $0.995126\pm0.000014$~d by calculating a Lomb-Scargle periodogram of the full dataset and fitting the peak of the power spectrum. Our value is within $1.5\sigma$ of the value from the DR24 release ($0.995144\pm0.000012$~d). Figure \ref{fig.phased} shows the light curve of \koinum\ phased with this period and centered on the transit. The transit is obvious in the phased data and the duration is about 4~hr. Both the best-fit (red curve) and a running median (grey curve) show this duration. This duration is much longer than expected  for a planet on a near-circular 1~d orbit around a 0.3\rsun{} star (45 minutes for impact parameter $b=0$). In the bottom panel of Fig. \ref{fig.phased} we show the phased data only for the last 600~d when the signal is stronger. The signal also has a 4~hr duration in this subset, therefore the long duration is not a product of time variation in the strength of the signal.

The duration and shape of the transit lightcurve constrain the density of the host star and orbital parameters (i.e. eccentricity) of the planet \citep[e.g.,][]{Seager2003}. A disparity between the expected vs. measured transit duration can indicate that (a) the signal is from a background or companion star with different properties, (b) the planet is on a highly eccentric orbit, and/or (c) the signal is not from a planet. To quantify this conflict, we compared stellar and orbital properties derived from the transits to those based on spectroscopy. 

We fit the stacked curves using a modified version of the Transit Analysis Package \citep[TAP,][]{Gazak2012}, which fits a \citet{MandelAgol2002} model with a quadratic limb-darkening law to the data. TAP employs 10 MC chains and records every 10th link in the chain. The model parameters are stellar density, limb-darkening (two for quadratic), planet-to-star radius ratio $R_P/R_*$, impact parameter $b$, orbital period ($P$), orbital eccentricity ($e$),  argument of periastron, epoch of first transit, and white noise. All but the first two are fit with uniform priors and bounded only by physical limitations (e.g., $-1.1<b<1.1$, $0<e<1$, $0<R_P/R_*<1$). 

As described in Gaidos et al. (2015), TAP was modified to calculate stellar density during each MCMC step following the formulae from \citet{Seager2003} so that a prior can be applied directly on stellar density rather than indirectly on transit duration. For limb-darkening we interpolated our stellar parameters (log~$g$, \teff, [Fe/H]) onto the \citet{Claret2011} grid of limb-darkening coefficients from the PHOENIX models, accounting for errors from the finite grid spacing, errors in stellar parameters, and variations from the method used to derive the coefficient (Least-Square or Flux Conservation); errors in both the linear and quadratic limb-darkening terms are $\simeq0.1$. 

\begin{figure}[]
\begin{center}
\includegraphics[width=\columnwidth]{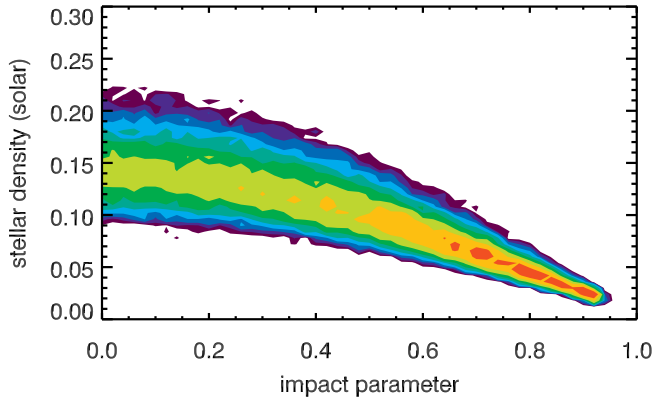}
\caption{Distribution of posterior values of the stellar density (in solar units) vs. the transit impact parameter from MCMC fit to the entire lightcurve dataset. Darker shades indicate a higher density of steps with those values. The period was set to the periodogram value and the eccentricity was set to zero; other parameters were allowed to float with no priors. \label{fig.nopriors}}
\end{center}
\end{figure}

We first fit the KOI~6705 lightcurve by fixing the period to the power spectrum value and the eccentricity to zero, with a prior on limb-darkening, but none on stellar density. Figure \ref{fig.nopriors} shows the distribution of the posterior values of the stellar density vs. the impact parameter $b$. The densities are all $\ll 1$ for any value of $b$, inconsistent with the expected value of $\approx 10$ based on spectroscopy (Sec. \ref{sec.star}).

Longer transit durations can occur if the planet's orbit is highly eccentric and the transit occurs near apapsis. To determine if this can reconcile the discrepant density estimates, we refit the light curve with eccentricity allowed to float, but applied a Gaussian penalty to the likelihood equivalent to the difference between transit-derived density and the spectroscopic (in standard deviations). For this we adopted a spectroscopic density with a mean of 9.5 and standard deviation of 2.0 in solar units (see Section~\ref{sec.star}).  

\begin{figure}[]
\begin{center}
\includegraphics[width=\columnwidth]{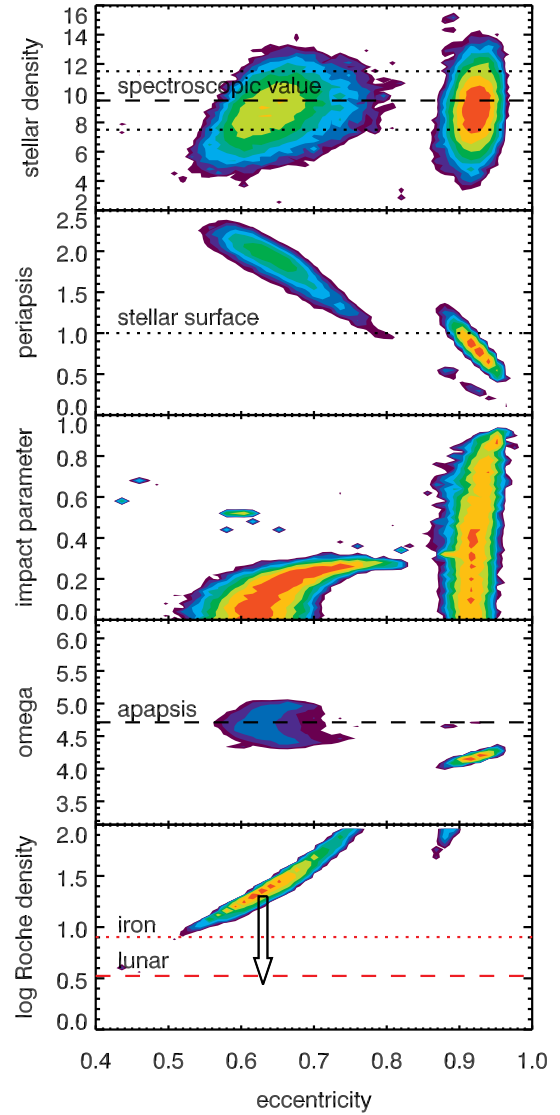}
\caption{Distribution of posterior values of the stellar density (in solar units), periapsis (stellar radii), impact parameter, argument of periapsis, and the minimum density which an incompressible, fluid-like body must have to avoid disruption by tides at periastron (common logarithm of values in g~cm$^{-3}$, see text). The color indicates the density of chains in parameter space, with red the highest. The color coding varies from plot to plot due to the differences in the concentration of links in parameter space. The dotted and dashed lines in the plot of stellar density mark the mean and plus/minus one standard deviation, respectively in the Gaussian prior function.  The dashed line in the periapsis plot is the stellar surface.  The dotted and dashed lines in the critical Roche density plot mark the density of pure uncompressed iron and the Moon, respectively.  The arrow indicates the displacement of minimum density to lower values if the body is rigid. \label{fig.priors}}
\end{center}
\end{figure}

Fig. \ref{fig.priors} plots the distributions of the posterior values of the stellar and orbital parameters for the MCMC chains, after removal of the "burn-in" and solutions where the stellar density is more than $4\sigma$ discrepant from the spectroscopic values.  Statistical values from this second set are reported in Table \ref{tab.params}.  Also plotted is the critical Roche density of an incompressible object with zero tensile strength at periapsis $r_{per}$ \citep{Roche1849}:
\begin{equation}
\rho = \rho \left(r_{per}/2.44R_*\right)^{-3}
\end{equation}
As expected, TAP found solutions where the orbit is highly eccentric and the transit occurs when the planet is near apapsis, producing the long transit duration. Actually, two sets of solutions were found, one clearly unphysical since the periapsis lies within the star. The second set of solutions is physically possible but geologically implausible; to escape destruction by tidal forces at periapsis, the density of such an object must exceed pure iron (bottom panel of Fig. \ref{fig.priors}). If the body were to have significant cohesive strength, the density requirement would be reduced by up to a factor of $(2.44/1.26)^3$ (arrow in bottom panel). 

Finally, the transit depth -- but not the duration -- unambiguously increased with time (Figs. \ref{fig.power} and \ref{fig.phased}).  The signal was undetectable in the first several observing quarters, and then strengthens with later quarters until the end of the mission.  This is most uncharacteristic of a planet.  A planet on a "grazing" transiting orbit (impact parameter $\approx 1$) might gradually appear (or disappear) as its orbit was perturbed by another planet (See Sec. \ref{sec.discussion} for discussion of transit timing variation).  But a grazing orbit would have a transit duration that was much shorter than the nominal 45~min, not much longer.   The transit depth and planet radius derived from an analysis of the full data set (Table \ref{tab.params}) are {\it average} values of uncertain interpretation.

\begin{table}
\centering
\caption{Properties of the KOI~6705 System}\label{tab.params}
\begin{tabular}{ll}
\hline
\multicolumn{2}{c}{Host Star \kicnum}\\
Spectral type & M3.5V\\
Effective temperature (K) & $3327 \pm 60$\\ 
Metallicity [Fe/H] & $-0.08 \pm 0.09$\\
Radius (\rsun) & $0.305 \pm 0.030$\\
Luminosity ($M_K$) & $7.05 \pm 0.30$\\
Mass\footnote{Based on Delfosse et al. (2000) mass-luminosity relation} $M_*$ (\msun{}) & $0.277 \pm 0.048$\\
Distance\footnote{based on $K$ apparent magnitude} (pc) & $70 \pm 10$\\
\hline
\multicolumn{2}{c}{Candidate Planet KOI~6705.01}\footnote{Based on analysis of all data}\\
Orbital period (d) & $0.995126\pm0.000014$\\
Orbital inclination (deg.) & $85.8^{+2.9}_{-5.2}$\\
Impact parameter & $0.07^{+0.36}_{-0.43}$\\
Orbital eccentricity & $0.73^{+0.20}_{-0.13}$\\ 
Radius (\rearth{}) & $0.258^{+0.034}_{-0.029}$\rearth\\
Equilibrium temperature\footnote{Assumes Mercury albedo of 0.067 and efficient heat redistribution} (K) & 760\\
\hline
\end{tabular}
\end{table}

\section{Discussion}
\label{sec.discussion}

We have confirmed a significant 0.995~d periodic, planetary transit-like signal in the \kep{} lightcurve of the middle-aged, solar-metallicity M dwarf star KOI~6705/\kicnum.  The transit depth plus the spectroscopically determined properties of the star indicate that the body, if a planet, has a radius of about 0.26\rearth{}, the size of the Moon. However, the signal is anomalous in that the duration of the transit ($\approx 2.7$~hr) is at least 3-4 times longer than expected for a near-circular orbit around this star, and the strength of the signal varies with time, increasing from an undetectable level through the four years of the prime \kep{} mission.  We recovered the signal of \koinum\ from the PDCSAP data using a completely independent analysis, with parameters close to the \kep{} pipeline values. We also recovered the signal and its peculiar behavior with an analysis of the uncorrected Simple Aperture Photometry (SAP) data.  Thus this signal is not an artifact of the data analysis.

We tested whether the transient nature of the signal might be the product of variation in stellar or instrumental noise. For each of the 12 quarters of the lightcurve we calculated the total noise as a robust standard deviation \citep{Tukey1977} and compared the distribution of values against the aggregate data set using the Kolmogorov-Smirnov test. The average of the 12 values is 422 ppm and its RMS among the quarters is 36 ppm. Simulated Monte Carlo datasets, constructed by sampling the aggregate dataset with replacement, have a mean of 367 ppm and vary by only 6.6 ppm. There is thus a significant ($6\sigma$) but small variation in stellar plus instrumental noise between quarters.  However, there is no apparent trend of the overall noise level with quarter that could explain the behavior of \koinum:  there are both lower- and higher-noise quarters in the first two years of data. Only the last quarter (16) has a noise distribution that is unambiguously different from the aggregate lightcurve (K-S test $p = 8 \times 10^{-12}$); it also has the highest noise (513 ppm). The other quarters all have $p>0.01$ and most have $p>0.1$. 

We used high-resolution AO imaging in the near-infrared to rule out companions with $K < 17$ as close as 0.15" from the star, a projected separation corresponding to 10~AU, and $K < 20$ at $\sim 1$" arc-second separations. These limits exclude all possible late M, L and most T dwarf companions \citep{Dupuy2012}. Although we could not rule out a companion star with a projected separation of $<10$~AU, there is no indication of a second star in our ESI spectrum, and a lower luminosity (and denser) host star would only aggregate the conflict between the transit and spectroscopic estimates of stellar density. Using imaging, photometric, and astrometric constraints, we found a negligible ($<3 \times 10^{-5}$) posterior probability that there is an appropriate background star in the \kep{} photometry aperture that could produce the signal as an EB.  We ruled out several other false-positive scenarios involving EBs, i.e. by scattered light from stars that are nearby on the sky, antipodal reflection, and cross-talk between detector readouts.  However, we identified one star, KIC~6503213, with a location, brightness, photometry-based properties, and variability consistent with it being an EB and producing \koinum\ via inefficient charge transfer along the column.  

\koinum\ is unlikely to be a planet.  The long transit duration can only be explained if the object is on a highly eccentric orbit ($e \sim 0.7$) and the transit occurs near apapsis.   The periapsis of possible orbits then lies either within the star or so close that the object would be disrupted by tides unless it had a mean density much greater than uncompressed iron\footnote{The compression inside a Moon-size body will be small.} or significant tensile strength. The circularization timescale of this object is
\begin{equation}
\tau_{\rm circ} = \frac{2}{21 \Im(k_2)}\frac{M_p a^{13/2}}{G^{1/2}M_*^{3/2}R_p},
\end{equation}
where $\Im(k_2)$, the imaginary part of the complex Love number, is the dissipation factor and the only completely unknown parameter. \citet{Driscoll2015} estimated $\Im(K_2) \sim 10^{-3}$ in the Earth using a Maxwell rheology to reproduce the present tidal dissipation. For this value, $\tau_{\rm circ} \sim 3 \times 10^{7}$~yr, which means observing a highly eccentric orbit is unlikely. However, tidal dissipation of orbital energy would heat the interior of the body, perhaps to the melting point, at which point the dissipation efficiency falls and $\tau_{\rm circ}$ increases by many orders of magnitude. Regardless, this scenario does not explain the time variation in the transit signal.

A signal like \koinum\ could, in principle, be produced by a transient, large but optically-thin dust cloud orbiting close to the star.  On an $e \approx 0$ orbit, an occultation duration of $3.3 \pm 0.3$~hr, is equivalent to a physical size of $1.5\times 10^{6}$~km, or about $7R_*$.  Dust clouds from ``disintegrating'' planets have been invoked to explain the lightcurves of KIC~12557548b \citep{Rappaport2012}, KOI-2700b \citep{Rappaport2014}, and K2-22b \citep{Sanchis-Ojeda2015}, and dust clouds associated with {\it forming} planets could explain the "dips" in the lightcurves of very young stars \citep{Ansdell2015}.  The mass of dust required to produce the obscuration over the transit depth is small, $\sim 10^{12}$~kg for 1 $\mu$m grains. Assuming the dust is removed in a single orbit, the required production rate is a few $10^7$ kg~sec$^{-1}$. This is only one order of magnitude higher than the dust production of comet Hale-Bopp near its perihelion at 0.9~AU \citep{Jewitt1999}.  One difficulty with this interpretation is that such a dust cloud need not produce a planet-like transit profile.  The apparent infrared excess suggested by the WISE data is intriguing, but any dust cloud's contribution to that emission would be minuscule:  assuming a cross-section $50 \times$ that of the stellar disk, an optical depth of $10^{-4}$, an emitting temperature of 700~K, and the Rayleigh-Jeans relation between spectral intensity and temperature, the relative flux from the dust cloud would be $\sim 1 \times 10^{-3}$, much smaller than the uncertainties in the WISE measurements. 

Instead, \koinum\ appears to be a FP produced by charge transfer inefficiency along the detector column that includes both \kicnum\ and a 1.99~d EB that falls 16.5' away.   Such column anomalies intensified, i.e. the FPs signals grew stronger, throughout the duration of the \kep{} mission, perhaps due to radiation damage to the CCDs (J. Coughlin, private communication).  Column anomaly would also explain the change in the power spectrum of the lightcurve of \kicnum\ during the latter two years of the mission (Fig. \ref{fig.rotation}).  The intensifying 45-50~d periodic signal may have arisen from the rotational variability of {\it another} star, e.g. any of the other three suspects we identified.

Our study of \koinum\ illustrates the care with which the smallest candidate planet signals must be analyzed and interpreted, especially at short periods where EBs are rampant, even for a well-behaved, well-characterized telescope like \kep{}.  Solving this mystery depended on the availability of FFIs obtained during transits/eclipses, a fortunate circumstance brought about by the long duration of the events compared to the orbital period.  The signals from most other \kep{} candidates have much shorter duty cycles and are not amenable to this approach.  Analysis of data from the Transiting Exoplanet Survey Satellite (TESS), which will survey most of the sky with 21" pixels, will encounter many of the same issues as \kep{} \citep{Sullivan2015}, but will benefit from the availability of full-frame images acquired at a 30-minute cadence; these should permit better localization of the source of any signal. 

\acknowledgements

The authors are indebted to Jeff Coughlin for his expert review and suggestion to use the \kep{} FFIs to evaluate false positive scenarios.  EG was supported by the International Short Visit program of the Swiss National Science Foundation.  This paper includes data collected by the \kep{} mission, funding for which is provided by the NASA Science Mission directorate. Some of the data presented herein were obtained at the W. M. Keck Observatory, which is operated as a scientific partnership among the California Institute of Technology, the University of California and NASA. The Observatory was made possible by the generous financial support of the W.M. Keck Foundation. This research has made use of the NASA Exoplanet Archive, which is operated by the California Institute of Technology, under contract with the National Aeronautics and Space Administration under the Exoplanet Exploration Program. This research has also made use of the Aladin Sky Atlas and the SIMBAD database, developed and operated at the CDS, Strasbourg Observatory in France.  

{\it Facilities:} \facility{IRTF:SpeX}, \facility{UH2.2m:SNIFS}, \facility{Keck2:ESI/NIRC-2}

\clearpage

\bibliographystyle{apj}

\clearpage

\end{document}